\documentclass[12pt]{iopart}
\usepackage{iopams}  

\usepackage{lineno,hyperref}
\usepackage{graphicx}
\newcommand {\bfv}[1] {\mbox{\boldmath ${#1}$}}
\newcommand\SEC[1]{\vskip 1mm {\it #1}.\rule[1mm]{5mm}{0.1mm}}

\begin{document}

\title{On the Formation of Unstirred Layer in Osmotically Driven Flow}

\author{Tomoaki Itano$^1$, Keito Konno$^2$, Taishi Inagaki$^1$, Masako Sugihara-Seki$^{1,3}$}
\ead{itano@kansai-u.ac.jp}
\address{
  $1$
  Department of Pure and Applied Physics,
  Faculty of Engineering Science, Kansai University,
  Yamatemachi 3-3-31, Suita, Osaka, 564-8680, Japan
}
\address{
  $2$
  Hitachi Solutions, Ltd.,
  Tokyo, 140-0002, Japan
}
\address{
  $3$
  School of Engineering Science, Osaka University, 
  Machikaneyama 1-3, Toyonaka, Osaka, 560-8531, Japan
}

\vspace{10pt}
\begin{indented}
\item[]August 2018
\end{indented}

\begin{abstract}
  Osmotically driven flow across a semi-permeable membrane under a constant static pressure difference is revisited with referring to the previous reports for reverse osmosis\cite{nak67,liu70}.
  A few mathematical techniques for obtaining the approximate solution, such as that for inverse problems used in the field of heat transfer, are presented with an emphasis on the nonlinear boundary condition and the time-dependent solvent flow-rate.
  It is concluded that the layer is spontaneously formed by osmosis rapidly in the time scaled by $\sim {\rm O}\bigl(\sqrt{t}\bigr)$, and that the layer thickness grows with no upper limit in an infinite time interval.
  Based on the obtained solution, we will also discuss the thermodynamical output work in an irreversible process which is extracted from the system as an osmotic engine.
\end{abstract}


%
\vspace{2pc}

\large
\renewcommand{\baselinestretch}{2.0}

\section{Introduction}
Osmosis in membrane transport is a physical mechanism underlying a variety of engineering or biological phenomena\cite{knu02}.
In desalination process\cite{phi16,aki18}, which is nowadays the essential requisite for supplying water to a rising population globally, fresh water is produced by filtration through a polymeric membrane by a large external hydrostatic pressure applied against ``reverse'' osmosis.
The membrane must suffers not only from hydraulic viscous resistance but also from the osmotic pressure depending on the solute concentration on the membrane.
Moreover, beyond an engineering expectation, the highly concentrated solution is accumulated in a thin boundary layer near the membrane at the high pressure side, which induces an unfavourable virtual resistance on the water filtration and eventually gives rise to a membrane fouling, which is of great importance for the recent engineering application.

On the other hand, ``forward'' osmosis means the spontaneous flow through a membrane mainly driven by concentration difference across the membrane subject to a small external hydrostatic pressure difference, which leads to the concentration relaxation attributed to the physics principle, {\it increase of entropy}.
The forward osmosis is practically utilised in the wide range of all living creatures, for example, cells and organs, such as epithelia of blood capillary or intestinal membrane in animals, or the translocation of nutrition in phloem network distributed in the whole body of plants\cite{jen09}, where any mechanical pumping system such as a heart in animals is absent.
The forward osmosis has recently attracted great attention as a new type of power generation or energy recovery system for future engineering innovation\cite{tza06,wan10}.

\begin{figure}[h]
  \begin{center}
    \includegraphics[angle=0,width=0.63\textwidth]{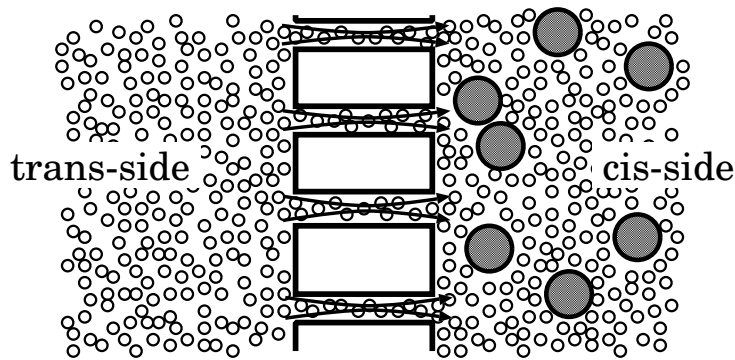}
  \end{center}
  \caption{Microscopic schematic view of forward osmosis process. Solvent and solute molecules are indicated by smaller and larger circles, respectively. 
The semi-permeable membrane locates at the center of figure as a sequence of rectangle regions.
The membrane prevents only solute molecules from leaking out through the membrane from the cis-side to the trans-side.
This steric barrier effect provides an uni-directional momentum selectively to solute molecules bombering the membrane by their Brownian motion, which leads to transmembrane volume flow of solvent\cite{mur93,ita08b,lio12}.
}
  \label{config}
\end{figure}

\SEC{Osmotically driven flow}
Here, we focus on {\it osmotically driven flow}, an ideal configuration of the forward osmosis so as to measure how quickly the osmosis spontaneously drives the net fluid movement under no external pressure difference across a membrane.
Suppose that initially pure solvent is partitioned from a solution of non-electrolyte with bulk concentration, $c_\infty$, by a semi-permeable membrane, for example, at the centre of a U-tube, which is used often to demonstrate the osmotic driven flow for science education.
From the microscopic viewpoint, steric barrier effect of the membrane selectively allows only solvent molecule passage but prevents solute molecules from leaking out through membrane pores (Fig.\ref{config}).

The solvent spontaneously starts to seepage across the membrane from the pure solvent side (trans-side) into the solution side (cis-side) due to the {\it osmotic pressure}, which virtually originates to the Brownian motion of molecules from the microscopic viewpoint\cite{mur93,ita08b,lio12}.
This transmembrane volume flux of solvent from trans-side to cis-side is sustained unless the equilibrium is achieved where the static hydraulic pressure difference across the membrane cancels the osmotic pressure.
According to classical thermodynamics, the osmotic pressure $\tilde{\Pi}$ of the ideal solution is deduced from the balance of the chemical potentials across the membrane, and is proportional to the product of the temperature and the solute concentration difference between cis- and trans-sides, which is known as van't Hoff law.
However note that, in reality, osmotic pressure, which is the strength of the macroscopic opposite pressure required to prevent the solvent seepage to the cis-side, is not a stress through a bulk fluid in a mechanical sense.
Thermodynamics tells us only the relaxation degree and direction to the equilibrium in terms of free energy, but not the time scale and solvent flow velocity in the primary transient process.
The seepage is a non-equilibrium and irreversible process which is still be one of challenging themes in modern physics, and we have to invoke the time-dependent continuum mechanics as well as the classical thermodynamics.

As time elapses without increase of opposing hydraulic pressure, the resultant transmembrane volume flux of solvent advects away solutes from the vicinity of the semi-permeable membrane, so that a certain layer of less solute concentration will be formed locally on the membrane in cis-side.
The developing layer spatially localised on the semi-permeable membrane has been named as either {\it concentration boundary layer}\cite{cur76,wan10} or as {\it unstirred layer}\cite{dai66,ped80}.
The layer could be the origin of the aforementioned fouling in the forward osmosis process as well as in the reverse osmosis.
In the forward osmosis, the decrease of solute concentration difference across membrane is inevitably followed by the reduction of the effective osmotic pressure and the transmembrane volume flux, although the development of the layer may be suppressed either by mechanical mixing with crossflow or by supply of solute molecules in the bulk region in cis-side.

\SEC{Objective}
The unstirred layer has been accounted to be one of unpredictable or unobservable factors of engineering problems appeared in desalination process, and thus its thickness is implicitly assumed to be saturated.
Actually, the presence of the unstirred layer is apparently plausible in practical cases, partly because the steady equilibrium has been mainly pursued in engineering viewpoint, and partly because the thickness of the unstirred layer is too thin to be observed in experiments and otherwise disturbed by an external flow brought by some bulk cross-flow tangential to the membrane.
Therefore, the detail discussion on the formation time-scale or the developoing thickness of the layer have hardly appeared in the previous literature with a few exceptions\cite{nak67,liu70}; how quickly does it develops ? Does its thickness saturate within a finite timescale ? 
In the present study, taking into account the unsteadiness of flow and concentration next to the membrane, we will revisit the problem to estimate analytically the formation timescale and thickness of the layer, by solving time-dependent governing equation of layer in the aforementioned ideal configuration.

\section{Formulation}
In what follows, the solute (number) concentration and flux and the solvent flow velocity are represented as $\tilde{c}$, $\tilde{\bfv{j}}$ and $\tilde{\bfv{u}}$, respectively, where an tilde symbol indicates a dimensional variable.
Hereafter, we shall restrict our attention to the case that these quantities are uniform along the membrane from the macroscopic point of view, so that they depend only on the time, $\tilde{t}$, and the distance from the membrane surface in cis-side, $\tilde{x}$.
We neglect the effect of gravity for ease of discussion, so that the natural convection due to the inhomogeneity of density distribution does not occur.
The macroscopic mass conservation rule is written by $\displaystyle \frac{\partial \tilde{c}}{\partial \tilde{t}}  + \tilde{\bfv{\nabla}}\bfv{\cdot}{\tilde{\bfv{j}}}=0$.
Following Fick's law, the flux $\tilde{\bfv{j}}$ is given approximately by the sum of advection and diffusion, $\tilde{\bfv{j}}=\tilde{c}\tilde{\bfv{u}}-\tilde{D}\tilde{\bfv{\nabla}} \tilde{c}$, where $\tilde{D}$ is the solute diffusivity in solvent.
The incompressibility of solvent flow may allow us to presume that $\tilde{\bfv{u}}$ is independent of $\tilde{x}$.
We adopt a couple of boundary conditions, $\tilde{\bfv{j}}\times \bfv{n}=0$ (the impermeability of the membrane with the normal vector $\bfv{n}$ against solutes at $\tilde{x}=0$) and $\tilde{c}=\tilde{c}_\infty$ (the uniform bulk concentration at $\tilde{x} \to \infty$).
These conditions mean no solute is supplied $\tilde{x}>0$.
An initial condition will be required for a fully posed initial-value problem.

\SEC{Negligible inertia}
If the solute concentration is so dilute, van't Hoff law is valid, $\tilde{\Pi}=\tilde{c}_0 \tilde{k}_{\rm B} \tilde{T}$, where $\tilde{\Pi}$ and $\tilde{c}_0$ are the osmotic pressure and  the concentration in the layer in contact with the membrane in the cis-side, $\tilde{c}(\tilde{t},\tilde{x}=+0)$, respectively.
It is empirically known that the solvent seepage velocity (volumetric flux) is proportional to the sum of the osmotic pressure and the static hydraulic pressure difference $\tilde{p}_0$ across the membrane, under no electrostatic potential difference across the membrane\cite{aki08}.
This proportionality is known as the classic Starling principle of fluid exchange, $\tilde{\bfv{u}}\cdot\bfv{n}=\sigma \tilde{L}_{\rm p} (\tilde{\Pi}-\tilde{p}_0)$, where in the present study the reflection coefficient of the membrane for solutes, $\sigma$, is unity and the hydraulic conductance (osmotic permeability, filtration coefficient) of the membrane, $\tilde{L}_{\rm p}$, is supposed to be a constant independent of concentration or property of impermeable solute.
The reference of $\tilde{p}_0$ is taken at trans-side.

In general, the hydraulic pressure difference $\tilde{p}_0$ may be a time-dependent variable, for example, a function of the waterlevel at time $t$, which is calculated as $\int_0^{\tilde{t}} \tilde{u}(\tilde{\tau}) d\tilde{\tau}$ in case of U-tube.
Hereafter, $\tilde{p}_0$ is kept to steady, and is zero unless otherwise noted.
Starling's relation may be practically satisfied in most of experiments, because the relaxation time-constant calculated from fluid inertia and hydraulic conductance is, in the most of cases, too small to be observable compared to the experimental timescale.
We can estimate the relaxation time-constant $\tilde{t}_{\rm relax}=\tilde{M} \tilde{L}_{\rm p}/\tilde{S}$ from the total fluid mass, $\tilde{M}$, and the membrane area, $\tilde{S}$.
Although these variables vary, of course, largely depending on the membrane thickness as well as material and size of solvent molecule, the magnitude of hydraulic conductance have been investigated experimentally for a variety of membranes, for example, ${\rm O}(10^{-11})$[m/s/Pa] for frog mesentery, ${\rm O}(10^{-12})$[m/s/Pa] for plasmalemma of Nitella translucens or Visking-dialysis tubing, ${\rm O}(10^{-13}) $[m/s/Pa] for toad skin in the literature\cite{cur76,esc72}.
Even if we roughly overestimate fluid mass and membrane area, we find that $\tilde{M}={\rm O}(1)$[kg], $\tilde{S}={\rm O}(1)$[mm$^2$], the time-constant is of the order of $1$[$\mu$s] at most.
Thus the inertia is negligible in a practical sense, so we hereafter presume that the solvent velocity is determined by the Starling's relation.

\SEC{Nondimensional equation}
Assembling all the aforementioned equations, via nondimensionalising time and length ($\frac{\tilde{t}}{\tilde{t}_0}\to t$, $\frac{\tilde{x}}{\tilde{x}_0}\to  x$) and via rescaling of dependent variable ($\frac{\tilde{c}}{\tilde{c}_\infty} \to  c$), we end up with the following second order nonlinear partial differential equation:
\begin{equation}
  \frac{\partial c}{\partial t} + (c_0-p_0) \frac{\partial c}{\partial x} = \frac{\partial^2 c}{\partial x^2}  \label{eq1}
\end{equation}
with a couple of boundary conditions,
\begin{equation}
 \lim_{x\to\infty} c= 1  \ \ \ \mbox{and}\  \ \   \frac{\partial c}{\partial x}\bigr|_{x=+0}=(c_0-p_0) c_0  \label{eq2}
\end{equation}
where $c_0$ is defined by $c_0 = c(t,x=+0)$, and $p_0=\tilde{p}_0/(\tilde{c}_\infty \tilde{k}_{\rm B} \tilde{T})$.
The nonlinearity in the advection term and the boundary conditions acts an flavour of difficulty in the problem.
The units of time and length are taken as $\tilde{t}_0=\tilde{D}/(\tilde{L}_{\rm p} \tilde{k}_{\rm B} \tilde{T} \tilde{c}_\infty)^2$, $\tilde{x}_0=\tilde{D}/(\tilde{L}_{\rm p} \tilde{k}_{\rm B} \tilde{T} \tilde{c}_\infty)$, respectively, so that the coefficient of advection and diffusion terms are both unity in the present study.
This nondimensionalisation was previously introduced by Nakano\cite{nak67} or Liu\cite{liu70}, although their focus was mainly in the reverse osmosis.
The similar but different nondimensionalisation has been adopted in the literatures\cite{ped78,ped80}, where nondimensionalisation are determined by some representative length or velocity units given by external conditions in individual cases, such as constant thickness of unstirred layer or hydraulic pressure difference across the membrane.
It should be noted that the present nondimensionalisation is the only possible way in our problem with $p_0=0$, where no external stirring factor is involved, that is, neither time nor length scale are given.
For a variety of aqueous solutions, $\tilde{D}$ is in the order of ${\rm O}(10^{-9})$[m$^2$/s], so that the units can be estimated as $\tilde{t}_0={\rm O}(
10^{-3}\sim 10^{+7} 
)$[s] and $\tilde{x}_0={\rm O}(
10^{-6}\sim 10^{-1} 
)$[m], respectively, in the reasonable ranges of $\tilde{c}_\infty \tilde{k}_{\rm B} \tilde{T} \sim {\rm O}(10^{5} \sim 10^{7}
)$[Pa] at room temperature ($\tilde{c}_\infty/N_{\rm A}={\rm O}(
10^{2}\sim 10^{4}
)$[Osmol/m$^3$]).

\section{Analysis}
\SEC{pseudo steady solution}
The obtained deterministic equation is difficult to be analytically integrated with a uniform initial condition, $ c(t=0)=1 $.
In the literature, the steady equilibrium observable in experiments has been mainly pursued, which would be realised independently of initial conditions uncontrollable in experiments.
Following Ref.\cite{dai66,sch74}, let us consider the case that the thickness of the unstirred layer in the equilibrium, $\vartheta$, saturates to a constant within a finite time interval.
Eliminating the time derivative term, we obtain 
$\displaystyle   \frac{\partial c}{\partial x} - (c_0-p_0) c  = {\rm Const.} $,
via integration of Eq.(\ref{eq1}) in $x$.
This condition can be satisfied by the following expression regardless of discontinuity of the first order spatial differential,
\begin{equation}
 c(x)= 1+\Theta(\vartheta-x)\bigl( c_0\exp{\bigl((c_0-p_0) x\bigr)}-1\bigr) \ \  \label{eq3}
\end{equation}
where $\Theta$ is the Heaviside step function.
The steady concentration profile is steepest at $x=\vartheta$, which is implausible in reality as previously pointed out by Ref.\cite{ped80}.
Although the magnitude of $\vartheta$ is determined by the relation $c_0={\rm e}^{-c_0 \vartheta}$ deduced from the boundary condition at $x=+0$, the arbitrariness of $c_0$ is still remained, which should be determined by introducing an external mixing effect in the bulk region into account.
On the other hand, in the present deterministic system without any external mixing in the bulk region, the magnitude of $c_0$ should be determined self-consistently.
As pointed out by Liu \cite{liu70}, who have conducted experimental observation of the reverse osmosis, the time-dependency is essential to study the present problem.

\begin{figure}[h]
  \begin{center}
    \includegraphics[angle=0,width=0.65\textwidth]{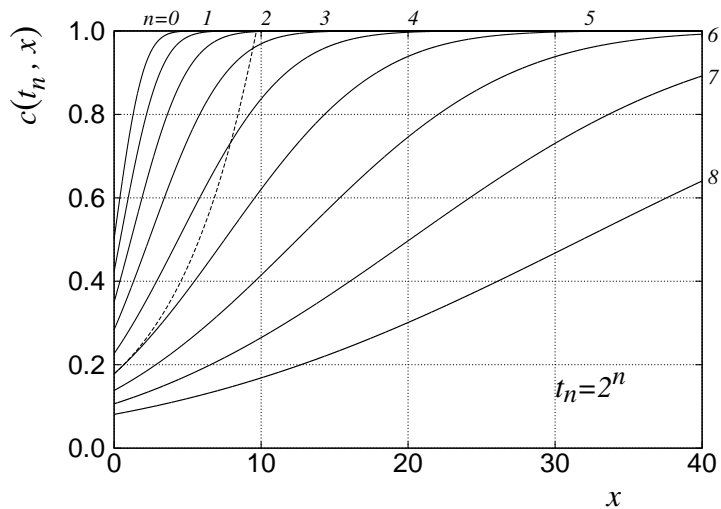}
  \end{center}
  \caption{Time development of nondimensionalised concentration distribution in solution side (cis-side, $x>0$) is numerically obtained using finite-difference method for the ideal case of $p_0=0$. Nondimensionalised time and length are $t=\tilde{t}/\tilde{t}_0$ and $x=\tilde{x}/\tilde{x}_0$, respectively, where $\tilde{t}_0=\tilde{D}/(\tilde{L}_{\rm p}\tilde{k}_{\rm B}\tilde{c}_{\infty}\tilde{T})^2$ and $\tilde{x}_0=\tilde{D}/(\tilde{L}_{\rm p}\tilde{k}_{\rm B}\tilde{c}_{\infty}\tilde{T})$. The unstirred layer, where concentration is less than that in the bulk region, is established in a relatively short time, $t \le t_0$, and the front of the layer proceeds far away from the membrane with increase of time.
  Dashed curve corresponds to the pseudo steady exact solution satisfying both conditions, $c(t,\vartheta)=1$ and $c_0={\rm e}^{-c_0 \vartheta}$ in case of $c_0=c(t_5,0)$.}
  \label{dns}
\end{figure}

\SEC{Numerical solution by finite-difference method}
For the purpose to verify that the steady state is not realised in a strict sense, we examined the numerical time-integration of the original time-dependent equation with the aforementioned initial condition and boundary conditions.
Fig.\ref{dns} shows the time series of nondimensionalised concentration profile, which is numerically integrated under forth-order Runge-Kutta scheme.
For the reference, $c(t_n,x)>0.999$ is satisfied for $x>4.62$ at $n=0$ and for $x>101$ at $n=8$, which implies that an artificial boundary at $x=200$ given in the simulation affects little to the result for a relatively early stage, $n\le 8$.
While the concentration in the layer in contact with the membrane, $c_0(t)$, is reduced to the half of unity abruptly within one time unit ($\tilde{t}/\tilde{t}_0 \le 1$), the decreasing rate of $c_0$ in time remarkably slows down over the time unit ($\tilde{t}\tilde{t}_0 \ge 1$).
In our system where no length scale exists except for $\tilde{x}_0$, the steady state does not seemed to be realised.
As time elapses further, the front of the unstirred layer proceeds far away from the membrane and the thickness of the layer increases asymptotically with no upper limit, which means the unstirred layer is not a boundary layer but can be thicker than literally would be envisaged\cite{jen09}.
The finiteness of characteristic thickness of the layer for the reverse osmosis was comprehensively discussed in the appendix of Ref.\cite{liu70}.
In other words, the unsteadiness of the system is related to the fact that the thickness of the unstirred layer is intrinsically undetermined.
However, of course, a finite length of the chamber or a finite measuring time interval in experimental restriction would make a steady state to be apparently realised.

For comparison, the pseudo steady profile satisfying both conditions, Eq.(\ref{eq3})$c_0={\rm e}^{-c_0 \vartheta}$ in case of $c_0=c(t_5,0)$, is additionally indicated by a dashed curve in the figure.
The discrepancy between the dashed curve and the solid curve of $n=5$ implies that the solute concentration on the membrane can be exaggerated (overestimated) by a simple extrapolation based on the macroscopic concentration measured at a distance far from the membrane under the assumption of steady equilibrium.

\SEC{Polynomial profile approximation}
The boundary condition at $x=+0$ is not satisfied by the initial condition.
The above numerical demonstration suggests that the unstirred layer develops in a relatively short time interval so as to regularise the singularity on the membrane rapidly.
In order to refine the above expression of the concentration, suppose that the thickness of the layer, $\vartheta$, and the local concentration on the membrane in cis-side, $c_0$, both are dependent on time, $t$, instead of Eq.(\ref{eq3}).
Moreover we assume that the profile of concentration, $c(t,x)$, is expressed by the $N$-th order polynomials of $x$ ($N\ge 1$) in the layer in contact with the membrane, $0<x<\vartheta(t)$ :
\begin{equation}
  c(t,x)= 1+\Theta(\vartheta(t)-x) \sum_{n=0}^N r_n(t) \bigl(\vartheta(t)-x\bigr)^n \ \ .\label{eq4}
\end{equation}

\begin{figure}[h]
  \begin{center}
    \includegraphics[angle=0,width=0.65\textwidth]{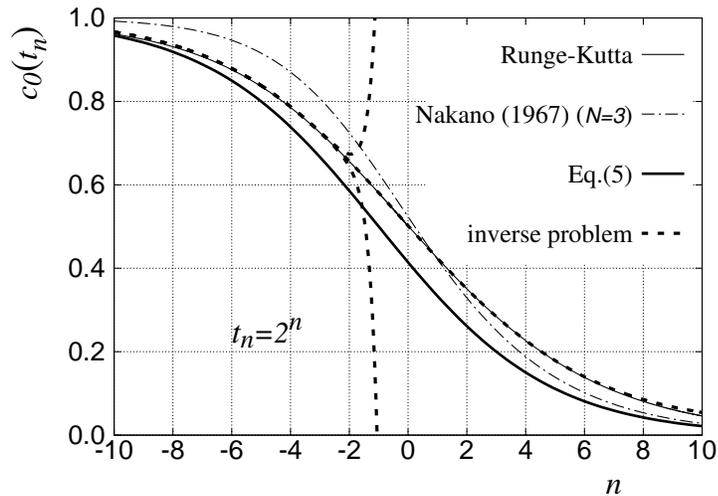}
  \end{center}
  \caption{Comparison of the nondimensionalised concentration at the vicinity of the membrane surface, $c_0$, between (thin solid curve) numerical solution and (thick solid curve) polynomial profile approximation at $N\to \infty$, Eq.(\ref{eq5}), for the ideal case of $p_0=0$.
    The intermidiate thick dashed curve designates the analytic continuation of the expression obtained from a procedure of inverse problem, which shows an excellent agreement with numerical result.
    The upper and lower dashed curves are equivalent to $c_0(t)$ in Eq.\ref{eq6} with the different truncation number $10$ and $11$.
  }
  \label{comp1}
\end{figure}

This expression is a refined expression of Eq.(\ref{eq3}).
Furthermore, we will assume the profile to be continuous and smooth enough at $x=\vartheta$, if saying in a strict sense, of differentiability class {\rm C}$^{N-1}$.
Thus, we obtain $r_n=(c_0(t)-1)\vartheta(t)^{-N}$ for $n=N$; otherwise $r_n=0$ ($0\le n \le N-1$).
Imposing that the governing equation is satisfied at $x=0$, we can deduce a first-order ordinary differential equation for $c_0(t)$,
\[ \frac{d c_0}{dt}= -\frac{N-c_0-(1-c_0)p_0}{(1-c_0)N}c_0^3 \ \ ,\]
which can be integrated analytically for any $N$, and thus leads to
\[
 -\frac{N-1}{N^2}\ln{\bigl(\frac{N-c_0}{(N-1)c_0}\bigr)}+\frac{1-c_0}{2c_0^2}(1-c_0+2\frac{c_0}{N}) = t \ \ ,
\]
for $p_0=0$.
From the boundary condition at the origin, we can deduce $\displaystyle \vartheta=\frac{N(1-c_0)}{(c_0-p_0)c_0}$, that is, the thickness of the layer, $\vartheta(t)$, will be later determined from $c_0(t)$.

The present problem originates in the quadratic second-order differential equation, Eq.(\ref{eq1}), so one would be interested specially in the case of $N=3$, under applying the continuity conditions at $x=\vartheta(t)$ for the first and second spatial derivatives and $c$ itself for an arbitrary $t$.
The polynomial profile approximation $c_0(t_n)$ obtained for a finite $N$ provides a qualitatively plausible time series of the concentration at the vicinity of the membrane surface, but quantatively distinct from the numerical result $c_0(t_0)=0.50$; for example, $c_0(t_0) \approx 0.44$ for $N=3$.

Moreover, one would be interested in the fact that the $N$-th order polynomial form addressed above converges to the simple expression at the limit $N\to \infty$, 
\begin{equation}
  c_0(t)=\frac{1}{1+\sqrt{2t}} \ \ \label{eq5}
\end{equation}
 in spite of $\vartheta\to\infty$.
Taking into account the original definition of exponential function, $\lim_{N\to\infty}\bigl(1+\alpha/N\bigr)={\rm e}^\alpha$, one will obtain the limit of the profile of the concentration at time $t$,
\[
  c(t,x)=1-(1-c_0(t))\exp{\bigl(-\frac{c_0(t)^2 x}{1-c_0(t)}\bigr)} \ \ .
\]
In the previous studies, the accurate measurement of the solute concentration at the vicinity of the membrane is challenged with the aid of the extrapolation of concentration.
As a measure of the thickness of the layer in the literature, $\Delta x$ is defined by $\displaystyle \frac{\partial c}{\partial x}|_{x=0} \cdot \Delta x = 1-c_0$.
In the present limit,  $\Delta x$ is also the function of time, $\displaystyle \Delta x=\sqrt{2t}+2t$.
Fig.\ref{comp1} shows the time series of the nondimensionalised solute concentration $c_0(t)$ at the origin given by Eq.(\ref{eq5}), compared to the numerical solution.

The comparison between the numerical result and Eq.(\ref{eq5}) in the figure suggests that a quantitative agreement is limited within the initial stage of the layer formation, $t\ll t_0$.
However, we would like to note that Eq.(\ref{eq5}) may be a lower bound of the numerical solution at least.
Taking into account that the nondimesionalised solvent velocity equals to $c_0-p_0$, the solvent volume seepaging for an infinite time interval may be calculated as $\int_0^{\infty} u(t) dt \sim \int_0^{\infty} \frac{1}{1+2t} dt \to \infty$.
Additionally, we note that the curve of $c_0(t_n)$ obtained from a finite $N$ exists between the numerical simulation and the limit case of $N\to \infty$, for example, $c_0(t_0) \approx 0.44$ for $N=3$, and $0.46$ for $N=2$ (not shown in the figure).

Nakano {\it et al} presented an analogous approach using a polynomial expression in the context of the reverse osmosis in Ref.\cite{nak67,liu70}, where the authors approximate the concentration profile as a cubic polynomial with the time-varying thickness of boundary layer.
Substituting the polynomial expression Eq.(\ref{eq4}) with $N=3$ into the integration of Eq.(\ref{eq1}) in $0<x<\vartheta(t)$, they obtained
\[ \frac{\partial}{\partial t} \int_0^{\vartheta(t)} c(t,x) dx + (c_0-p_0)=0 \ \ ,\]
so as to deduce another first-order ordinary differential equation for $c_0(t)$, instead of imposing that the governing equation is satisfied at $x=0$.
Following their procedure, we obtain the ordinary equation for an arbitrary $N$, 
\[ \frac{d}{dt}\Bigl(\frac{(1-c_0)^2}{(c_0-p_0)c_0}\Bigr) = \frac{N+1}{N}(c_0-p_0) \ \ ,\]
and the implicit solution for $p_0=0$ is
\[ \frac{1}{3}\Bigl(\frac{1-c_0}{c_0}\Bigr)^3 + \frac{1}{2}\Bigl(\frac{1-c_0}{c_0}\Bigr)^2 = \frac{N+1}{2N} t  \ \ .\]
For comparison, the above implicit expression of $c_0$ on $t$ for $N=3$ is indicated as the dash-dotted curve in Fig.\ref{comp1}.

\begin{figure}[h]
  \begin{center}
    \includegraphics[angle=0,width=0.55\textwidth]{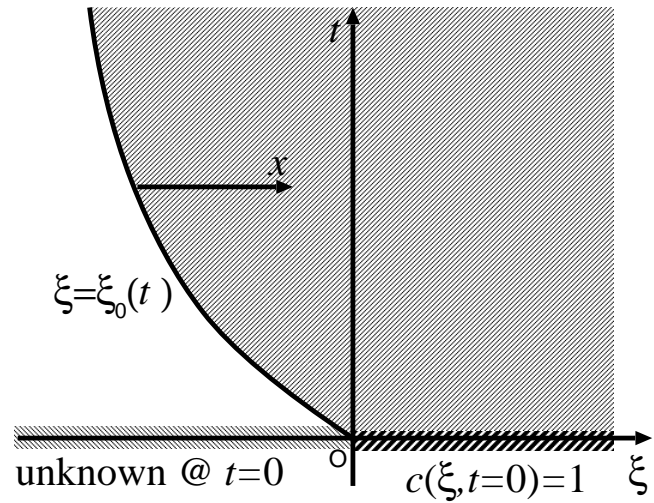}
  \end{center}
  \caption{
Schematic view of $t-\xi$ plane used in the present inverse problem. The shaded region in $\xi>\xi_0(t)$ corresponds to cis-side. The unknown initial value on the half-line $\xi<0$ at $t=0$ is solved so as to satisfy the boundary condition Eq.(\ref{eq2}) on the curve $\xi=\xi_0(t)$.
  }
  \label{fig_schematic}
\end{figure}

\SEC{Approach in inverse problem}
Finally, we will apply a procedure in the inverse problem to the present problem, Eq.(\ref{eq1}) and Eq.(\ref{eq2}).
With leaving $c_0(t)$ unsolved provisionally, we transform the frame of reference $x \to \xi$, where $x=\xi-\xi_0(t)$ and $\xi_0(t)$ is determined by
\begin{equation}
  \frac{d\xi_0}{dt}=-c_0(t)+p_0 \ \ \ .   \label{eq6}
\end{equation}
By the coordinate transformation, Eq.(\ref{eq1}) is converted into the standard linear diffusion equation in the $\xi-t$ space ($-\infty<\xi<\infty$ and $0<t<\infty$), which is solvable analytically for a well-posed initial condition in terms of appropriate Green function.
It should be however noted that the initial condition in $\xi\le0$ is arbitrary for the moment, while $c(t,\xi)\bigr|_{t=0}=1$ is given for $\xi > 0$(see Fig.(\ref{fig_schematic})).
Thus, for example, we may assume the initial condition expanded in the $-\infty<\xi<\infty$ is represented as
\[ 
c(t,\xi)\bigr|_{t=0}= 1+ \Theta(-\xi) \sum_{n=1}^{\infty} a_n \xi^n\ \ ,
\]
in terms unknown coefficients, $a_n$, so that $\displaystyle \lim_{\xi \to -0} c(t,\xi)=\lim_{\xi \to +0} c(t,\xi)$ is satisfied at $t=0$.
From the initial condition, the exact solution $c(t,\xi)$ for $t>0$ is determined as 
\begin{equation}
  c(t,\xi)=1+\sum_{n=1}^{\infty} a_n I_n(\tau,\eta) = c(\tau,\eta) \ \ ,\label{eqIn}
\end{equation}
where $\displaystyle (\tau,\eta)=(\sqrt{t},\frac{\xi}{\sqrt{4t}})$, and
\begin{eqnarray*}
  I_0(\tau,\eta)&=&\frac{1}{2} {\rm erfc}(\eta) \ \ ,\\
  I_1(\tau,\eta)&=&2 \tau \eta I_0(\tau,\eta)-\frac{\tau}{\sqrt{\pi}}\exp{(-\eta^2)} \ ,\\
  I_n(\tau,\eta)&=&2 \tau \eta I_{n-1}(\tau,\eta)+2(n-1)\tau^2I_{n-2}(\tau,\eta) \ \ \ \ \mbox{for} \ \ \ n\ge 2\ \ .
\end{eqnarray*}
By the way, taking into account that $c_0(t)$ is a function of $\sqrt{t}$ in the approximation discussed before (see Eq.(\ref{eq5})), we may suppose that $\xi_0(t)$ determined by Eq.(\ref{eq6}) is expanded into a series as
$\displaystyle \xi_0(t) = \sum_{n=0}^\infty b_n t^{\frac{n}{2}}$, where $b_0=b_1=0$ from $\xi_0(0)=0$ and $b_2=-1$ from $\displaystyle \frac{{d\xi}_0}{dt}(0)=-c(0,0)+p_0=-1+p_0$.
Suppose $\displaystyle \eta_0(\tau)=\frac{\xi_0(t)}{\sqrt{4 t}}=\frac{1}{2}\sum_{n=2}^{\infty} b_n \tau^{n-1}$, then
Eq.(\ref{eq6}) is converted to
\begin{equation}
  c\bigl(\tau,\eta_0(\tau)\bigr)-p_0+  \frac{\eta_0(\tau)}{\tau}+\frac{d \eta_0}{d\tau} =0\ \ \ , \label{eq7}
\end{equation}
which will provide a family of restriction on unknown coefficients of the series, $a_n$ ($n\ge 1$) and $b_n$ ($n\ge 3$).
The remained boundary condition corresponding to Eq.(\ref{eq2}), which should be satisfied at $\xi=\xi_0(t)$, is
\begin{equation}
  c\bigl(\tau,\eta_0(\tau)\bigr) \Bigl\{ c\bigl(\tau,\eta_0(\tau)\bigr)-p_0 \Bigr\} =  \frac{1}{2\tau} \frac{\partial c}{\partial \eta} \bigr|_{\tau,\eta_0(\tau)} \label{eq8}
\end{equation}
in the $(\tau,\eta)$ space. 
Eq.(\ref{eq7}) and Eq.(\ref{eq8}) subject to Eq.(\ref{eqIn}) constitute a set of algebraic recurrence relations for unknown coefficients of the series, $(a_n,b_n)$.
The procedure may be classified into the so-called inverse problem, which does not guarantees whether the solution converges properly, in general.
The coefficients for the case of $p_0=0$ can be determined as
\begin{eqnarray*}
  &&
  a_1=2 \ \ , \ 
  a_2=\frac{5}{2} \ \ , \ 
  a_3=\frac{33\pi+32}{18\pi} \ \ , \ 
  a_4=\frac{819\pi+896}{576\pi} \ \ ,\ \cdots\\
  &&
  b_3=\frac{4}{3\sqrt{\pi}} \ \ , \ 
  b_4=-\frac{3}{4} \ \ ,\ 
  b_5=\frac{27\pi+128}{45\sqrt{\pi}^3} \ \ , \ 
  b_6=-\frac{339\pi-128}{288\pi}\ \ , \ \cdots \ .
\end{eqnarray*}
It is numerically confirmed that the obtained form can be convergent only within the relatively small radius of convergence.
The convergence may be envisaged from the separation point around $(n,c_0)=(-2,0.65)$, where the upper and lower thick dashed curves obtained by the truncation numbers $10$ and $11$ detaches each other in Fig.(\ref{comp1}).
Following a mathematical procedure proposed in Ref.\cite{tak84}, we furthermore perform an analytic continuation by
\begin{equation}
c_0(t) = \sum_{n=0}^\infty q_n \bigl( \frac{\sqrt{t}}{\sqrt{t}+1} \bigr) ^n \ \ \ , \label{eq9}
\end{equation}
so as to improve the rate of convergence of the obtained alternating series.
In fig.\ref{comp1}, the intermidiate dashed curve designates the expression obtained from a procedure of inverse problem, which provides an excellent agreement with the numerically integratated solution.
\begin{table}[h]
  \centering
  \begin{tabular}[h]{|l|l|l|l|}
  \hline  \hline
  $n$ & $a_n$ & $b_n$ & $q_n$ \\  \hline  \hline
  0 & 0      & \ 0      & \ 1 \\  \hline
  1 & 2      & \ 0      &-1.1284 \\  \hline
  2 & 2.5    &-1      &-0.3716 \\  \hline
  3 & 2.3992 & \ 0.7523 &-0.2517 \\  \hline
  4 & 1.9170 &-0.75   & \ 0.1084 \\  \hline
  5 & 1.3319 & \ 0.8493 &-0.0806 \\  \hline
  6 & 0.8267 &-1.0356 & \ 0.0337 \\  \hline
  7 & 0.4670 & \ 1.3256 &-0.0291 \\  \hline
  8 & 0.2432 &-1.7561 & \ 0.0101 \\  \hline
  9 & 0.1180 & \ 2.3872 &-0.0115 \\  \hline
 10 & 0.0538 &-3.3104 & \ 0.0025 \\ \hline
  \end{tabular}
  \caption{
    Coefficient of the series $(a_n,b_n,q_n)$ solved in case of $p_0=0$. The convergence of $a_n$ and $b_n$ is not excellent and provides the small radius of convergence. While $b_n$ and $c_n$ are alternating series, $a_n$ is also potentially alternating series because $a_n$ is defined in Eq.(\ref{eqIn}) originating at algebraic series of $\xi^n$ for $\xi<0$.
  }
  \label{table1}
\end{table}

\section{Concluding remarks}
In the present study, we revisited a nonlinear partial differential equation describing the osmotically driven flow across a semi-permeable membrane under a constant static pressure difference, which was previously investigated for reverse osmosis\cite{nak67,liu70}.
Firstly we confirmed the three points in the unstirred layer formed in osmotically driven flows; (1) inertia is practically negligible, (2) layer formation timescale of the order of ${\rm O}(\tilde{t}_0)$ may be variable depending the parameters, $1$[ms]$\sim 1$[year],  depending on the individual case studied (3) the layer thickness may grow with no limitation in principle.
Following the previous literatures, the thickness of the layer, $\vartheta$, has been evaluated by the equation, $c_0={\rm e}^{(p_0-c_0)\vartheta}$, which is deduced under the assumption of pseudo steady solution of Eq.(\ref{eq1}). 
In the present study, keeping in mind that the phenomenon is essentially time-dependent, we applied a few mathematical procedures to the present problem for the purpose to measure the timescale for concentration boundary layer on the membrane to develop.
Specially, the last procedure for the inverse problem is successful to provide an analytical expression of the time series of solute concentration, Eq.(\ref{eq9}) with Table \ref{table1} for $p_0=0$ case.
From the result, we found that the localised concentration on the membrane decaying initially as $-t^{\frac{1}{2}}$ and eventually as $t^{-\frac{1}{2}}$ (cf. \cite{vil81,don95}).
This implies that the unstirred layer thickness $\vartheta$ diverges with increase of $t$ with no upper limit and the layer intrinsically never saturates.

Here, we will refer the present system to an osmotic engine that can convert osmotic energy to mechanical work. 
The solvent volume seepaged into cis-side is calculated as $|\xi_{p_0}(t)|=\int_0^{t}\bigl( c_0(\tau)-p_0 \bigr) d\tau$ per an area on the membrane.
The nondimensionalised work output $\delta W$ for the initial time interval $t$ is a function of $t$ and $p_0$, which equals to the product $p_0$ and $|\xi_{p_0}(t)|$ because the present seepages is an isobaric process in thermodynamics.
Note that the process under the condition $p_0=0$ corresponds to the irreversible free expansion, and results only in increase of entropy (no work output $\delta W=0$), which may complete in a relatively short time.
On the other hand, the flow under the condition $p_0=0$ corresponds to the reversible and quasistatic process, which requires a longer time interval to be completed.
From these facts, we may conjecture that the maximum power $\delta W/t$ is realised at a optimal nondimesionalised pressure $p_0$ ($0<p_0<1$).
Based on Eq.(\ref{eq6}) and $c_0(0)=1$, we obtain for a small $t$,
\begin{eqnarray*}
\delta W &=& \frac{dW}{dt}\bigr|_{t=0} t + \frac{1}{2}\frac{dW}{dt}\bigr|_{t=0} t^2+ {\rm O}(t^3)  \\
&\approx& \Bigr( -\bigl(p_0-\frac{1}{2}\bigr)^2 + \frac{1}{4}\Bigr) t + \frac{1}{2}p_0 \frac{d c_0}{dt}\bigr|_{t=0} t^2 \ \ .
\end{eqnarray*}
From the equation, we can conclude that the initially instantaneous output power is the maximum $0.25$ at $p_0=0.5$, and that the average output power for a longer $t$ is less than 0.25 because $\frac{\partial c_0}{\partial t}\bigr|_{t=0}<0$.
The integration of Eq.(\ref{eq9}) based on the numerically obtained $c_n$ shows that $\delta W/t$ decreases to $0.144$ at $t=1$.
Moreover, we can predict that the optimal value of $p_0$ for a longer $t$ is less than $0.5$ because $\displaystyle \frac{\partial^2 c_0}{\partial p_0 \partial t} \Bigr|_{t=0}<0$.

Finally we note that the initial condition in the present study is not artificial.
For instance, the solvent flow can be initially prevented by the static pressure equal to the osmotic pressure, $p_0=c_\infty$, then the initial condition in the present study is easily realised by extingushing $p_0$.
In physiology, it is known that such a sudden change of the solute concentration (or the static pressure) gives rise to a certain dysfunction on the membranes of cells.
With the aid of some mechanics, cells or tissue in our body are able to respond such a osmotic shock in the short formation timescale of the unstirred layer neutralizing the influence.



\section{Acknowledgement}
We thank to 
Dr T. Ooshida, Prof H. Isozaki and Prof. N. Sugimoto for fruitful discussion at an earlier draft of the manuscript.
T.I. is grateful for the financial support in part by the Kansai University Subsidy for Supporting Young Scholars, 2017.

\bibliographystyle{iopart-num}

\bibliography{osmosis01}
\end{document}